\documentclass[reqno,11pt]{article}  
\newcommand{\version}{June 11,  2008}
\newlength{\dinwidth}
\newlength{\dinmargin}
\usepackage{amsmath}
\usepackage{amssymb}
\usepackage{latexsym}
\usepackage{cite}
\usepackage{epsfig,psfrag}
\newcommand{\RR}{\mathbb{R}}

\newcommand{\CC}{\mathbb{C}}

\newcommand{\unity}{{\setlength{\unitlength}{1em}
                     \begin{picture}(0.75,1)
                     \put(0,0){$1$}
                     \put(0.34,0){\line(0,1){0.65}}
                     \end{picture}
                   }}

\newcommand{\Ad}{\text{\rm Ad} }
\newenvironment{Proof}%
{\par \medskip \noindent {\em Proof.}}{\hspace*{\fill} $\square$\par%
\medskip\noindent}
\newtheorem{Thm}{Theorem}
\newtheorem{Prop}{Proposition}
\newtheorem{Lem}{Lemma}

\newcommand{\calM}{{\mathcal M}}

\newcommand{\calH}{{\mathcal H}}
\newcommand{\calA}{\mathcal{A}}          
          
\newcommand{\calW}{\mathcal{W}}   
\newcommand{\calO}{\mathcal{O}}

\newcommand{\Po}{P_+^{\uparrow}}
\newcommand{\Poj}{P_+}

\newcommand{\clo}{ {\mbox{\bf --}} }

\newcommand{\spc}{{C}}             
\newcommand{\Spc}{\mathcal{K}}         
 
\newcommand{\Boox}[1]{\lambda_{#1}}           
\newcommand{\J}{j}         
\newcommand{\Auni}{\calA} 
\newcommand{\AO}{\calA_0}
\newcommand{\HO}{\calH_{0}}
\newcommand{\Om}{\Omega}

\newcommand{\lsp}{(\,}
\newcommand{\rsp}{\,)}

\newcommand{\barrho}{{\bar{\rho}}} 
 
\newcommand{\DO}{\Delta_0}  
\newcommand{\JO}{J_0}  
\newcommand{\SO}{S_0}  %
\newcommand{\UO}{U_0}   
\newcommand{\alphaO}{\alpha}   

\newcommand{\CPT}{CPT }

\newcommand{\coc}{Z}

\begin{document} 
\title{Borchers' Commutation Relations for Sectors with Braid Group
  Statistics in Low Dimensions}
\author{Jens Mund\thanks{Supported by FAPEMIG.}
\\ 
\scriptsize 
Departamento de F\'{\i}sica, Universidade Federal de Juiz de Fora,\\  
\scriptsize 
36036-900 Juiz de Fora, MG, Brazil.\\ 
\scriptsize  E-mail: {\tt mund@fisica.ufjf.br}}
\date{\version 
}
\maketitle 
\begin{abstract}
Borchers has shown that in a translation covariant vacuum
representation of a theory of local observables with positive energy
the following holds: The (Tomita) modular 
objects associated with the observable algebra of a fixed wedge 
region give rise to a representation of the subgroup of the Poincar\'e
group generated by the boosts and the reflection associated to the
wedge, and the translations. 
We prove here that Borchers' theorem also holds in charged sectors with 
(possibly non-Abelian) braid group statistics in 
low space-time dimensions. Our result is a crucial step towards 
the Bisognano-Wichmann theorem for Plektons in $d=3$, 
namely that the mentioned modular objects 
generate a representation of the proper Poincar\'e group, 
including a \CPT operator.  
Our main assumptions are Haag duality of the observable algebra, and 
translation covariance with positive energy as well as finite
statistics of the sector under consideration.   
\end{abstract}
\section*{Introduction} \label{secIntro}
Borchers has shown~\cite{Borchers92} that in a theory of local
observables, which is translation covariant with positive energy, 
the modular objects associated with the observable algebra of a
(Rindler) wedge region and the vacuum state have certain specific 
commutation relations with the representers of the translations. 
Namely, these commutation relations manifest that 
the corresponding unitary modular group implements the 
group of boosts which leave the wedge invariant, and that the corresponding 
modular conjugation implements the reflection about the edge of the wedge. 
Borchers' theorem has profound consequences. For example in two-dimensional 
theories it means that the modular objects generate a representation
of the proper Poincar\'e group, under which the observables behave covariant, 
and implies the \CPT theorem. 
In higher dimensions, it is a crucial step towards the Bisognano-Wichmann
theorem in the general context of
local quantum physics~\cite{Borchers98,BY00,BGL93,GL00,Kuck00,SHW98,M01a}. 
This theorem asserts that a certain class of Poincar\'e covariant
theories enjoys the property of modular covariance, namely that the
mentioned unitary modular group coincides with the 
representers of the
boosts, and that the modular conjugation is a \CPT operator (where
`PT' means  the reflection about the edge of the wedge). 

The hypothesis under which Borchers' theorem works is the double role 
played by the vacuum vector within a theory of local algebras: The
vacuum is cyclic and separating for the local algebras, and it is  
invariant under the positive energy representation 
of the translation group under which these algebras are covariant. 
In a charged sector, i.e., a non-vacuum representation of the
observables, this situation is not given. (This problem has been posed
by Borchers 
in~\cite[Sect.\ VII.4]{Borchers2000}.) In the case of permutation group
statistics, one can use the field algebra instead of the observable
algebra to recover the result.  
However, in low-dimensional space-time there may occur
superselection sectors with braid group statistics~\cite{F89,FM1}. 
Then only in the Abelian case there is a field (C$^*$) algebra
for which the vacuum is cyclic and separating. In the case of
non-Abelian braid group statistics, there is no such field algebra. 
Due to this complication, a general result corresponding to Borchers' 
theorem has not been achieved yet.   
In the present article, we prove an analogue of Borchers'
theorem for a superselection sector corresponding to a localizable 
charge. 
The implementers of the boosts and the reflection which we find are 
the relative modular objects associated with the observable algebra of
the wedge, the vacuum state and some specific state in the conjugate sector.   
We assume that the observable algebra satisfies Haag duality, see 
Eq.~\eqref{eqHD}, and that the sector under consideration has
finite statistics and positive energy, and is irreducible.   
We also need a slightly stronger irreducibility
property~\eqref{eqRhoIrred}, which may be 
ensured by requiring for example Lorentz covariance or the split
property. 
We consider charges which are localizable in space-like 
cones\footnote{A space-like cone is, in $d\geq 3$,  
a convex cone in Minkowski space generated by a double cone and a point 
  in its causal complement, and in $d=2$ the causal completion
  thereof, which is a wedge region.}, 
and admit the case of non-Abelian braid group statistics which can
occur in low space-time dimensions, $d=2$ and $3$. 

It must be noted that in two dimensions, our result is already
practically covered by the work of Guido and
Longo~\cite{GL92}. Namely, they show how a certain condition of
modular covariance in the vacuum sector allows, under the
same hypothesis as in the present article, for the construction of a 
(ray) representation of the proper Poincar\'e group in charged
sectors. But in two dimensions, their modular covariance condition is
satisfied due to Borchers' theorem (in the vacuum sector), so their
analysis goes through, even in sectors with Braid group statistics.  
However, it must be noted that in $d=2$ the assumption of Haag duality 
excludes some massive models with braid group statistics as e.g.\ the
anyonic sectors of the CAR algebra~\cite{Adler}, and together with the
split property for wedges (expected to hold in massive models) excludes
localizable charges altogether~\cite{Mueger98}.  

Our result shall be used to derive the \CPT and Bisognano-Wichmann theorems for
particles with braid group statistics in three-dimensional
space-time~\cite{Mu_BiWiAny}. 
It would be gratifying to extend our analysis to soliton sectors in 
2 dimensions, which would extend the range (and simplify the proof) of
Rehrens' \CPT theorem for solitons~\cite{Re98}. 
\section{General Setting, Assumptions and Results} \label{Ass}
We consider a theory of local observables, given by a family of 
von Neumann algebras $\AO(\calO)$ of operators acting in the 
vacuum Hilbert space $\HO$, indexed by the double cones 
$\calO$ in Minkowski space, and satisfying the conditions of isotony
and locality: 
\begin{align*} 
\AO(\calO_1) \subset \AO(\calO_2)\;\text{ if }\; \calO_1\subset
\calO_2\quad \text{ and } \quad  
\AO(\calO_1) \subset \AO(\calO_2)' \;\text{ if }\; \calO_1\subset
\calO_2' ,
\end{align*}
where the prime denotes the commutant or the causal complement, respectively. 
The vacuum Hilbert space $\calH_0$ carries a unitary 
representation $\UO$ of the group of space-time translations $\RR^d$ with
positive energy, i.e.\ its spectrum\footnote{By spectrum of a
  representation of the translation group we 
mean the energy-momentum spectrum, namely the joint 
spectrum of the generators.} lies in the forward light cone. 
It has a unique, up to a phase, invariant vector $\Omega\in\HO$,
corresponding to the vacuum state. 
The representation $\UO$ implements 
automorphisms 
under which the net $\calO\to \AO(\calO)$ is covariant: 
\begin{align} 
\Ad \UO(x)\,\AO(\calO) &= \AO(x+\calO) \label{eqCovObs} 
\end{align}
for all $ x\in\RR^d$. (By $\Ad U$ we denote the adjoint action of a
unitary $U$.) 

Borchers' theorem, which we wish to generalize to charged sectors,
asserts that the representation $\UO$ has specific commutation
relations with certain algebraic objects, the so-called modular
group and conjugation, which suggest a geometric interpretation 
of the latter.  
Let us recall Borchers' commutation relations in this setting. 
Let $W_1$ be the wedge defined as  
\begin{align} \label{eqWe}  
W_1 & := \{\,x\in\RR^d:\,|x^0|<x^1\;\}\,. 
\end{align} 
By the Reeh-Schlieder property, $\Omega$ is cyclic and 
separating for the von Neumann algebra $\AO(W_1)$ generated by all 
$\AO(\calO)$, $\calO\subset W_1$.   
This allows for the definition of the Tomita operator~\cite{BraRob}, $\SO$,  
associated to $\AO(W_1)$:  
It is the  closed anti-linear involution satisfying 
\begin{align}  \label{eqSTomO}
\SO\,A\Omega&=A^*\Omega, \quad A\in\AO(W_1)\,.
\end{align}
Its polar decomposition, 
$\SO = \JO\,\DO^{1/2}$,  
defines an  anti-unitary involution $\JO$, the so-called {modular conjugation},
and a positive operator $\DO$ giving rise to the so-called 
{modular unitary group} $\DO^{it}$ associated to the wedge $W_1$. 
By Tomita's Theorem, see e.g.~\cite{BraRob}, 
the adjoint action of $\DO^{it}$ leaves $\AO(W_1)$ invariant, 
and the adjoint action of $\JO$ maps $\AO(W_1)$
onto its commutant $\AO(W_1)'$. 
The mentioned theorem of Borchers now asserts that $\DO^{it}$ and
$\JO$, together with the representation $\UO$ of the translations, 
induce a representation of the subgroup of
$\Poj$ generated by the boosts $\Boox{t}$ and the 
reflection $j$ associated to the wedge, and the translations. More precisely, 
let $\Boox{t}$ be the (rescaled) 1-boosts, leaving $W_1$ invariant and
acting on the coordinates $x^0,x^1$ as  
\begin{equation} \label{eqBoox}  
\left( \begin{array}{cc}
 \cosh(2\pi t) &  \sinh(-2\pi t)  \\
 \sinh(-2\pi t) &  \cosh(2\pi t)
 \end{array} \right), 
\end{equation} 
and let $j$ be the reflection about the edge of $W_1$, 
acting on the coordinates $x^0,x^1$ as $-\unity$ and leaving the other
coordinates unchanged (if $d>2$). 
Then Borchers' theorem asserts that 
\begin{align} 
\DO^{it}\,\UO(x)\,\DO^{-it}&= \UO(\Boox{t}x), \label{eqDUD0}\\
\JO\,\UO(x)\,\JO &= \UO(jx) \label{eqJUJ0}
\end{align}
for all $t\in\RR$ and $x\in\RR^d$.  
These relations implement the group relations 
of the translations with the boosts and reflections, respectively. 
Modular theory further implies that $\JO$ is an involution and
commutes with the modular unitary group, implementing the
group relations $\J^2=1$ and $\J\Boox{t}\J^{-1}=\Boox{t}$. 
Altogether, $U_0(x)$, $\DO^{it}$ and $\JO$ constitute a 
representation of the subgroup of the Poincar\'e group generated 
by the translations, the boosts $\Boox{t}$ and the reflection $\J$ 
(which is the direct product of the proper Poincar\'e group in the
time-like $x^0$-$x^1$ plane and the translation group in the 
remaining $d-2$ dimensions).   

Our aim is to find a similar result in a charged sector, 
that is in a representation of the abstract $C^*$-algebra generated by
the local algebras $\AO(\calO)$, which is inequivalent from the defining vacuum
representation. 
We shall consider an irreducible representation $\pi$, which is 
localizable in space-like cones. That means that $\pi$ and
the vacuum representation are unitarily equivalent in restriction to
the  observable algebra associated with the causal complement of any space-like
cone.\footnote{It is known that every purely massive 
representation is localizable in space-like cones~\cite{BuF}.} 
We assume that the observable algebra satisfies Haag duality for 
space-like cones and wedges, i.e., regions which arise by a proper 
Poincar\'e transformation from $W_1$. Namely, denoting by $\Spc$ the class 
of space-like cones, their causal complements, and wedges, we require  
\begin{equation} \label{eqHD}
\AO(\spc')= \AO(\spc)',\quad \spc \in\Spc. 
\end{equation}
A localizable representation can then be described by an endomorphism of
the so-called universal algebra $\Auni$ generated by isomorphic
images $\Auni(\spc)$ of the $\AO(\spc)$, $\spc\in\Spc$, 
see~\cite{F90,GL92,FRSII}. 
The family of isomorphisms $\Auni(\spc)\cong \AO(\spc)$ extends to a 
representation $\pi_0$ of $\Auni$, the vacuum representation. We then have 
\begin{equation} \label{eqAO} 
\AO(\spc)= \pi_0\Auni(\spc), 
\end{equation}
and the vacuum representation is faithful\footnote{However, $\pi_0$
  is in general not faithful on the global algebra $\Auni$  
due to the  existence of global intertwiners~\cite{FRSII}.} 
 and normal on the local\footnote{We call the algebras
  $\Auni(\spc)$ ``local'' although the regions $\spc$ extend to
  infinity in some direction, just in distinction
  from the ``global'' algebra $\Auni$.} algebras
$\Auni(\spc)$. 
The adjoint action~\eqref{eqCovObs} of the translations on the local algebras 
lifts to a representation by automorphisms $\alphaO_x$: 
\begin{align} \label{eqU0impl} 
\Ad \UO(x) \circ \pi_0 &= \pi_0\circ \alphaO_x,\\
\alphaO_x\,  \Auni(\spc)&=   \Auni(x+\spc). \label{eqCovObs'} 
\end{align}
Our localizable representation $\pi$ is then
equivalent~\cite{DHRIII,F90} with a 
representation of the form $\pi_0\circ \rho$ acting in $\HO$, 
where $\rho$ is an endomorphism of $\Auni$ {localized} in some 
specific space-time region $\spc_0\in\Spc$ in the sense that 
\begin{equation} \label{eqRhoInC0}
\rho(A)=A\qquad \text{ if } \; A \in\Auni(C_0'). 
\end{equation} 
We shall take the localization region of $\rho$ to be properly contained in 
$W_1$, which implies by Haag duality~\eqref{eqHD} that $\rho$
restricts to an endomorphism of $\Auni(W_1)$. 
We shall require that this endomorphism of $\calA(W_1)$ be
irreducible, namely that  
\begin{equation} \label{eqRhoIrred}
\pi_0 \calA(W_1)\cap \big(\pi_0\rho\calA(W_1)\big)' = \CC \unity. 
\end{equation}
This is a slightly stronger requirement than irreducibility of the
representation $\pi_0\rho$ of $\Auni$. 
It has been shown by Guido and Longo that irreducibility of
$\pi_0\rho$, together with finite statistics, imply irreducibility in
the sense of Eq.~\eqref{eqRhoIrred} if $\rho$ is covariant under the 
(proper orthochronous) Poincar\'e
group~\cite[Cor.~2.10]{GL96}~\footnote{Although not explicitly mentioned 
in~\cite{GL96}, the proof does not depend on covariance of $\rho$
under the full Moebius group. See also~\cite[Thm.~2.2]{Longo96}.} 
or if $\rho$ satisfies the split 
property~~\cite[Prop.~6.3]{GL92}. 
We further assume the representation $\pi\cong \pi_0\rho$ to be
translation covariant with positive energy. That means that there is a 
unitary representation 
$U_\rho$ of the translation group $\RR^d$ with spectrum 
contained in the forward light cone such that 
\begin{equation} \label{eqCovRho} 
\Ad U_\rho(x) \circ \pi_0\rho = 
\pi_0\rho \circ \alphaO_x, \quad x\in\RR^d. 
\end{equation} 
We finally assume that $\rho$ has finite statistics, i.e.\ 
that the so-called 
statistics parameter $\lambda_\rho$~\cite{DHRIII} be non-zero. 
This holds automatically if $\rho$ is massive~\cite{F81}, and  
implies~\cite{DHRIV} the existence of a 
{\em conjugate} morphism $\bar\rho$ 
characterized, up to equivalence, by the fact that the composite 
sector $\pi_0\bar\rho\rho$ 
contains the vacuum representation $\pi_0$ precisely once.
Thus there is a unique, up to a factor, intertwiner $R_\rho\in\Auni(\spc_0)$
satisfying $\barrho\rho(A) R_\rho = R_\rho A$ for all $A\in\Auni$.  
The conjugate $\barrho$ shares with $\rho$ the properties of 
covariance~\eqref{eqCovRho}, finite statistics, and
localization~\eqref{eqRhoInC0} in
some space-like cone which we choose to be $\spc_0$. 
Using the normalization convention of \cite[Eq.~(3.14)]{DHRIV}, 
namely $R_\rho^*R_\rho=|\lambda_\rho|^{-1}\unity$, 
the positive linear endomorphism $\phi_\rho$ of $\Auni$  defined as
\begin{equation} \label{eqLeftInvR}
\phi_\rho(A)= |\lambda_\rho|\; 
R_\rho^*\bar\rho(A) R_\rho 
\end{equation}
is the unique left inverse~\cite{DHRIV,BuF} of $\rho$. 
In the low-dimensional situation, $d=2,3$, the statistics parameter 
$\lambda_\rho$ may be a complex non-real number, 
corresponding to braid group statistics. We admit the case when its 
modulus is different from one (namely when $\rho$ is not surjective), 
corresponding to non-Abelian braid group statistics. 
  
The modular objects for which we shall prove Borchers' commutation
relations are defined  as follows. 
Let $S_\rho$ be the  closed anti-linear operator satisfying 
\begin{align}  \label{eqSTomRho}
S_\rho \, \pi_0(A)\Om &:= \pi_0\barrho(A^*)R_\rho\Om, \quad 
A\in\Auni(W_1), 
\end{align} 
and denote the polar decomposition of $S_\rho$ by $S_\rho= 
J_\rho\Delta_\rho^{1/2}$. 
$S_\rho$ is just the relative Tomita operator~\cite{Stratila}  
with respect to a certain pair of (non-normalized) states. Namely,
consider the vacuum state $\omega_0:=\lsp \Om,\pi_0(\cdot)\Om\rsp$, 
and the positive functional 
$$
\varphi_\rho:= |\lambda_\rho|^{-1}\; \omega_0\circ \phi_\rho 
= \big(R_\rho\Om,\pi_0\bar\rho(\cdot) R_\rho\Om\big).  
$$
The restriction of $\varphi_\rho$ to $\Auni(W_1)$ is faithful and
normal, and has the 
GNS-triple $(\HO,\pi_0\barrho,R_\rho\Omega)$. 
Thus, $S_\rho$ is the relative Tomita operator associated with the
algebra $\Auni(W_1)$ and the pair of states $\omega_0$ and 
$\varphi_\rho$, see Appendix~\ref{secRelMod}.  
The motivation to consider these objects (instead of the modular
objects associated with $\Auni(W_1)$ and one 
suitable state, e.g.\ $\varphi_\rho$) is that the so-defined relative 
modular unitary 
group $\Delta_\rho^{it}$ implements the modular automorphism group 
associated with $\Auni(W_1)$ and $\omega_0$ in the
same way as the representation $U_\rho(x)$ implements the translations 
$\alphaO_x$, see Eq.~\eqref{eqDeltaImplement} below. 
This opens up the possibility to lift Borchers' commutation 
relations~\eqref{eqDUD0} in the vacuum representation to the 
representation $\pi_0\rho$. 
In fact, pursuing this strategy, we shall find the following result.  
Let $G$ be the subgroup of the proper Poincar\'e group generated by
the translations,
the boosts $\Boox{t}$ and the reflection $\J$. 
Recalling that the representation $U_\rho$ may be shifted to a 
representation $e^{ik\cdot x} U_\rho(x)$ whose spectrum has a Lorentz 
invariant lower boundary~\cite{BoBu85},\footnote{This is automatically
  the case if $\rho$ is localizable in double cones and $d>2$ by a result of
  Borchers~\cite{Borch65}, which is applicable since in this case 
  $\rho$ is implemented by local
  charged field operators~\cite{DR90}. It is also the case of 
  course if $U_\rho$ extends to the Poincar\'e group.} we show 
under the above-mentioned assumptions: 
\begin{Thm}[Commutation Relations.]  \label{Borchers}
Assume that the lower boundary of the spectrum of $U_\rho$ is
Lorentz-invariant. 
Then $U_\rho(x)$, $\Delta_\rho^{it}$, $J_\rho$ and the counterparts 
for $\barrho$ constitute a continuous (anti-) unitary 
representation\footnote{Strictly speaking, a ray representation since 
  $J_\rho J_\barrho$ is only a multiple of unity.} of
$G$. More specifically, there hold the commutation relations 
\begin{align} 
\Delta_\rho^{it}\,U_\rho(x)\,\Delta_\rho^{-it}&= U_\rho(\Boox{t}x), 
\label{eqDUD}\\
J_\rho\,U_\rho(x)\,J_\rho^{-1} &= U_\barrho(jx), \label{eqJUJ} \\
J_\rho\;\Delta_\rho^{it}\; J_\rho^{-1}&=\Delta_\barrho^{it}, \label{eqDJD} \\
J_\rho\,J_\barrho &=\chi_\rho \unity , \label{eqJJ} 
\end{align}
for all $t\in\RR$ and $x\in\RR^d$.  
The complex number  $\chi_\rho$ in Eq.~\eqref{eqJJ} has modulus one,
conjugate $\bar \chi_\rho=\chi_\barrho$ and is a root of unity 
if $\barrho=\rho$.  
\end{Thm}
(Note that Eq.~\eqref{eqDJD} corresponds to a standard property of 
{\em modular} objects, but needs to be proved for our relative modular 
objects.) 

We also show that this representation of $G$ acts geometrically
correctly on the wedge algebras, namely for $W$ in the family $\calW_1$ of 
translates of $W_1$ and $W_1'$, 
$$
\calW_1:=\{ x+W_1,\, x\in\RR^d\}\cup\{ x+W_1',\, x\in\RR^d\}, 
$$
there holds 
\begin{align} \label{eqDGeo}
\Ad \Delta_\rho^{it} :\,\pi_0\rho\Auni(W)&\to  \pi_0\rho \Auni(\Boox{t}W), \\
\Ad J_\rho           :\,\pi_0\rho\Auni(W) &\to
\pi_0\barrho\Auni(jW). \label{eqJGeo}  
\end{align}
To this end, observe that modular theory~\cite{BraRob} and the relation 
$\pi_0^{-1}(\AO(W_1)')=\Auni(W_1')$ imply that $\DO^{it}$ and $\JO$ 
implement an automorphism $\sigma_t$ of $\Auni(W_1)$ and
$\Auni(W_1')$, and an 
anti-isomorphism from $\Auni(W_1)$ onto $\Auni(W_1')$ and vice versa, 
respectively, defined by  
\begin{align}  \label{eqDelta0Sig} 
\Ad \DO^{it}\circ\pi_0&=\pi_0\circ\sigma_t \\
\Ad J_0\circ \pi_0 &= \pi_0\circ \alphaO_j\label{eqJ0alphaj} 
\end{align}
on $\Auni(W_1)\cup\Auni(W_1')$. By Borchers' commutation relations,
the same equations extend $\sigma_t$ and $\alphaO_j$ to the family $\Auni(W)$,
$W\in\calW_1$, acting in a geometrically correct way:
\begin{align} \label{eqD0Geo}
\sigma_t :\,\Auni(W)&\to  \Auni(\Boox{t}W), \\
\alphaO_j:\,\Auni(W)&\to \Auni(jW), \label{eqJ0Geo}  
\end{align}
$W\in\calW_1$, see~\cite[Lem.~III.2]{Borchers92}.   
But our representers $\Delta_\rho^{it}$ and $J_\rho$
implement these isomorphisms $\sigma_t$ and
$\alphaO_j$, respectively, in the direct product representation 
$\pi_0\rho\oplus\pi_0\barrho$, namely: 
\begin{Prop}[Implementation Properties.] \label{DeltaJImplement}
There holds 
\begin{align}\label{eqDeltaImplement}
\Ad \Delta_\rho^{it} \circ \pi_0\rho &=  \pi_0\rho \circ \sigma_t \\
\Ad J_\rho \circ \pi_0\rho & =\pi_0\barrho\circ \alphaO_j \label{eqJImplement}
\end{align}
on the family of algebras $\Auni(W)$, $W\in\calW_1$. 
\end{Prop}
Since $\sigma_t$ and $\alpha_j$ act geometrically correctly, 
c.f.~Eq.s~\eqref{eqD0Geo} and \eqref{eqJ0Geo}, this implies  that
$\Delta_\rho^{it}$ and $J_\rho$ act geometrically correctly, as 
claimed in Eq.s~\eqref{eqDGeo} and \eqref{eqJGeo}. 

In two space-time dimensions, our group $G$ 
already coincides with
the proper Poincar\'e group $\Poj$, and our results therefore imply that
the translations and the relative modular objects constitute an
(anti-) unitary representation of the latter.  
By our assumption of Haag duality~\eqref{eqHD} for wedges, the 
so-called dual net
$$ 
\Auni^d(\calO) :=  \bigcap_{W\supset \calO} \Auni(W) 
$$ 
is still local. (One needs to intersect in fact only the algebras of 
one ``right wedge'' of the form $W_1+x$ and one 
``left wedge'' of the form $W_1'+y$.) 
The modular (anti-) automorphisms $\sigma_t$ and 
$\alpha_j$ act on it in a geometrically correct way, 
see~\cite[Prop.~III.3]{Borchers92}. 
If the original net satisfies Haag duality also for double cones,
it coincides with the dual net. 
Then the implementation properties~\eqref{eqDeltaImplement} 
and \eqref{eqJImplement} hold, and therefore the 
representation of $\Poj$ constructed in Theorem~\ref{Borchers} acts 
geometrically correctly, namely there holds for any double cone $\calO$: 
\begin{align} 
\Ad U_\rho(g)&: \pi_0\rho\Auni(\calO) \to \pi_0\rho\Auni(g \,
\calO),\quad  g\in \Po, \nonumber\\
\Ad J_\rho&:\pi_0\rho\Auni(\calO)\to\pi_0\barrho\Auni(\J\, \calO).
\label{eqJRhoO}
\end{align}
Here we have written $U_\rho(a,\Boox{t}):=U_\rho(a)\Delta_\rho^{it}$. 
In  particular, $J_\rho$ is a \CPT operator.\footnote{If the net does
  not satisfy Haag duality for double cones, it does not coincide
 with the dual net. Then our endomorphism $\rho$ has two generally 
 distinct extensions $\rho_{R/L}$ to the dual net, according a choice
 of the right or left wedge~\cite{Roberts76}. (Each of them is
 localizable only in one type of wedges.) In this case, $J_\rho$
 intertwines $\pi_0\rho_R$ with $\pi_0\barrho_L\alphaO_j$, and 
 in Eq.~\eqref{eqJRhoO} there appears $\rho_R$ on one side and $\bar\rho_L$ on
the other side.}  
Again, it must be noted that these results (in $d=2$) are already
implicit in the work of Guido and Longo~\cite{GL92}, and also that the 
split property 
would exclude any charged sectors in our sense. 
\section{Proofs}
\label{secProof} 
We now prove Theorem~\ref{Borchers} and
Proposition~\ref{DeltaJImplement}. 
Instead of proving Borchers' commutation relations directly 
(e.g.\ paralleling Florig's nice
proof~\cite{Florig98}), we show how they lift from the
vacuum sector to our charged sector. 
We shall use some well-known facts about relative modular objects, 
which we recall in the Appendix for the convenience of the reader,  
see also~\cite{Stratila} for a review. 
Namely, the operator $\Delta_\rho^{it}\Delta_0^{-it}$ is in 
$\pi_0\Auni(W_1)$ for $t\in\RR$, and we define 
\begin{equation} \label{eqCoc}
\coc_\rho(t):= \pi_0^{-1}\big(\Delta_\rho^{it}\Delta_0^{-it}\big) 
\quad \in \Auni(W_1). 
\end{equation} 
This family of observables coincides with the 
Connes cocycle $(D\varphi_\rho:D\omega_0)_t$ with respect to the pair of 
weights $\omega_0$ and $\varphi_\rho$, see Eq.~\eqref{eqConnesCoc}.  
%
In the present context, it satisfies 
\begin{equation} \label{eqCocInt}
\Ad \coc_\rho(t)\circ \sigma_t \circ \rho = \rho \circ \sigma_t 
\qquad \text{ on } \calA(W_1), 
\end{equation}
see  Proposition 1.1 in~\cite{Longo97}.
The definition~\eqref{eqCoc} and Eq.~\eqref{eqCocInt}
are analogous to well-known properties of the translation cocycles
which we shall use in the sequel. Observe that for $a\in W_1^\clo$, the closure
of $W_1$, we have $W_1+a\subset W_1$ and $W_1'-a \subset
W_1'$. Since $\rho$ acts trivially on $W_1'$, this implies that 
the operator $U_\rho(a)\UO(-a)$ is in $\pi_0\Auni(W_1')'$ which
coincides with $\pi_0\Auni(W_1)$ by Haag duality. 
This gives rise to the translation cocycle 
\begin{equation} \label{eqCocY}
Y_\rho(a):= \pi_0^{-1}\big(U_\rho(a) \UO(-a)\big)\quad \in
\Auni(W_1),\; a\in W_1^\clo.  
\end{equation}
By virtue of Eq.s~\eqref{eqU0impl} and \eqref{eqCovRho}, it satisfies
the intertwiner relation 
\begin{equation} \label{eqCocYInt}
\Ad Y_\rho(x) \circ \alphaO_x \circ \rho = \rho \circ \alphaO_x,
\quad x\in \RR^d.
\end{equation} 
The definitions of the cocycles $\coc_\rho(t)$ and $Y_\rho(x)$, the
intertwiner relations~\eqref{eqCocInt} and \eqref{eqCocYInt}, and
invariance of $\Om$ under $\DO^{it}$ and $\UO(x)$ imply the 
identities 
\begin{align} \label{eqZpi0AOm}
\Delta_\rho^{it}\,\pi_0(A)\Om &= \pi_0\big(\coc_\rho(t)\sigma_t(A)\big)\Om,
\;\quad A\in \Auni(W_1),\\
U_\rho(x)\,\pi_0(A)\Om &= \pi_0\big(Y_\rho(x)\alphaO_x(A)\big)\Om,
\quad A\in \Auni, \label{eqYpi0AOm}
\end{align}
which we shall frequently use in the sequel. 
We shall also use the fact that Borchers' theorem applied to the
observable algebra implies that 
\begin{equation}\label{eqSigmaAlphaSigma}
\sigma_t \alphaO_{\Boox{-t}x} \sigma_{-t} = \alphaO_{x}
\end{equation}
holds as an isomorphism from $\Auni(W)$ onto $\Auni(W+x)$,  
$x\in\RR^d$, $W\in\calW_1$. 
Finally, we make the interesting observation that 
$S_\rho$ is the relative Tomita operator associated
not only with the pair of states $(\omega_0,\varphi_\rho)$, but
also with  pair of states $(\varphi_{\barrho},\omega_0)$: 
\begin{Lem} \label{SRhoOm}
The span, $D$,  of vectors of the form
$\pi_0[\rho(A)R_\barrho]\Omega$, $A\in\Auni(W_1)$, is
a core for the relative Tomita operator $S_\rho$, and $S_\rho$ acts
on $D$ as 
\begin{align}  \label{eqSTomRho'}
S_\rho\; \pi_0\rho(A)R_\barrho\Omega =  \chi_\rho \; \pi_0(A^*)\Omega, \quad 
A\in\Auni(W_1), 
\end{align} 
where $\chi_\rho$ is a complex number of modulus one, with 
$\bar\chi_\rho= \chi_\barrho$, and is a root of unity if $\barrho=\rho$.  
\end{Lem}
\begin{Proof}
Eq.s~\eqref{eqCocInt}, \eqref{eqZpi0AOm} and \eqref{eqZZRR} imply that for
$A\in\Auni(W_1)$ there holds 
$$
\Delta_\rho^{it}\;\pi_0[\rho(A)R_\barrho]\Omega = 
\pi_0\big[ \rho\sigma_t(A\coc_\barrho(-t))
 R_\barrho\big]\Omega. 
$$
Thus,  the domain $D$ is invariant under the unitary group 
$\Delta_\rho^{it}$. 
It is therefore a core for $\Delta_\rho^{1/2}$ and hence for $S_\rho$. 
On this core, we have by definition 
$$
S_\rho\; \pi_0[\rho(A)R_\barrho]\Omega=
\pi_0[\barrho(R_\barrho^*)R_\rho A^*]\Omega. 
$$
But $\barrho(R_\barrho^*)R_\rho$ is a self-intertwiner of $\rho$, hence a 
multiple of unity, $\chi_\rho\unity$. This proves Eq.~\eqref{eqSTomRho'}. 
For the stated properties of $\chi_\rho$, see~\cite[Eq.~(3.2)]{FRSII}.  
\end{Proof}
Since $(\HO,\pi_0\rho,R_\barrho\Omega)$ is the GNS triple for the
(non-normalized) state $\varphi_\barrho$ and $\chi_\rho\Omega$ is a
GNS vector for $\omega_0$, the Lemma implies that  
$S_\rho$ is the relative Tomita operator associated
with the pair of states $(\varphi_{\barrho},\omega_0)$. 
\paragraph{Proof of Theorem~\ref{Borchers}.} 
To prove Eq.~\eqref{eqDUD} of Theorem~\ref{Borchers}, let 
$A\in\Auni(W_1)$ and $a\in W_1^\clo$. 
Using Eq.s~\eqref{eqZpi0AOm}, \eqref{eqYpi0AOm} and 
\eqref{eqSigmaAlphaSigma}, we then have 
\begin{align} \label{eqDUDOm}
\Delta_\rho^{it} U_\rho(\Boox{-t}a)\Delta_\rho^{-it} \,\pi_0(A)\Om 
& \,= 
\pi_0\big(\hat Y_\rho(a,t) \alphaO_{a}(A)\big)\Om,  \\
\hat Y_\rho(a,t)& :=
\coc_\rho(t)\sigma_t\big(Y_\rho(\Boox{-t}a)\alphaO_{\Boox{-t}a} 
(\coc_\rho(-t)\big) . \nonumber
\end{align}
The intertwiner relations~\eqref{eqCocInt} and \eqref{eqCocYInt} imply
that 
on $\Auni(W_1)$ there holds 
\begin{align*} 
\Ad \hat Y_\rho(a,t) \circ \alphaO_{a} \circ \rho 
& \equiv 
\Ad \coc_\rho(t) \circ \sigma_t \circ \Ad Y_\rho(\Boox{-t}a) \circ 
\alphaO_{\Boox{-t}a} \circ \Ad \coc_\rho(-t) 
\circ \sigma_{-t}  \circ \rho \\
&= \rho \circ \sigma_t \circ \alphaO_{\Boox{-t}a} \circ \sigma_{-t} 
= \rho \circ \alphaO_a. 
\end{align*}
That is, $\hat Y_\rho(a,t)$ satisfies the same intertwiner 
relation~\eqref{eqCocYInt} on $\Auni(W_1)$ as $Y_\rho(a)$.  
On the other hand, $\hat Y_\rho(a,t)$ is also contained in
$\Auni(W_1)$. 
Therefore $\hat Y_\rho(a,t)Y_\rho(a)^*$ is in 
  $(\rho\Auni(W_1))'\cap\Auni(W_1)$ which is trivial by our
assumption~\eqref{eqRhoIrred} of irreducibility. 
Thus $\hat Y_\rho(a,t)$ coincides with 
$Y_\rho(a)$ up to a 
scalar function $c(a,t)$. Hence Eq.~\eqref{eqDUDOm} reads 
\begin{align*} 
\Delta_\rho^{it} U_\rho(\Boox{-t}a)\Delta_\rho^{-it} \,\pi_0(A)\Om 
& = c(a,t)\; \pi_0\big(Y_\rho(a) \alphaO_{a}(A)\big)\Om
\\
&\equiv  c(a,t)\; U_\rho(a) \, \pi_0(A)\Om. 
\end{align*}
Since the vacuum is cyclic for $\pi_0\Auni(W_1)$ by the Reeh-Schlieder 
property, this shows that 
\begin{align} \label{eqDUD'}
\Delta_\rho^{it} U_\rho(\Boox{-t}a)\Delta_\rho^{-it}=
c(a,t)\; U_\rho(a) 
\end{align}
for $a\in W_1^\clo$. By adjoining, we get an analogous equation for
$-a\in W_1^\clo$. Since the closures of $W_1$ and $-W_1$ span the
whole Minkowski space, this shows that there is a function $c(a,t)$
such that Eq.~\eqref{eqDUD'} holds for all $a\in \RR^d$. 
It remains to show that $c(a,t)\equiv 1$. 
Eq.~\eqref{eqDUD'} gives us a ray representation of the group  
$G$ generated by 
the boosts $\Boox{t}$ and the translations in the $0,1$-plane, defined by 
\begin{equation*} 
 U(a,\Boox{t}) := U_\rho(a)\Delta_\rho^{it}.
\end{equation*}
(The group $G$ is a subgroup of $\Po$ in $d=3$ and coincides with
$\Po$ in $d=2$. The product in $G$ is 
$
(a,\Boox{t})\cdot (a',\Boox{t'})= (a+\Boox{t}a',\Boox{t+t'})$.) 
Now $G$ is simply connected, and its second cohomology group is known
to be trivial. 
Therefore there exists a function $\nu$ from
$G$ into the unit circle such that 
$\hat U(g):= \nu(g)\,  U(g)$ is a true representation of $G$.    
Eq.~\eqref{eqDUD'} then implies that 
\begin{equation} \label{eqcnua}
c(a,t) = \nu(a,\unity) \,
\nu(\Boox{-t}a,\unity)^{-1}. 
\end{equation}
Since $U_\rho$ is a true representation of the translations, the 
restriction of $\nu$ to the
translations is a one-dimensional representation, that is of the form 
$\nu(a,\unity)=e^{ik\cdot a}$. 
Therefore, the spectra of the representations $\hat U=\nu\otimes
 U$ and $ U$ differ by a translation about a vector $k$. 
But the spectrum of $\hat U$ is invariant under the 1-boosts  
since $\hat U$ extends to a true 
representation of the (2-dimensional) Poincar\'e group $G$, and the
lower boundary of the spectrum of $ U$ is also Lorentz\ invariant 
since it coincides with
the spectrum of $U_\rho$. This implies that $k=0$ and hence, by
Eq.~\eqref{eqcnua}, that $c(a,t)\equiv 1$.  
This completes the proof of Eq.~\eqref{eqDUD} of the Theorem. 

We now prove Eq.~\eqref{eqDJD} of the Theorem. 
For $A\in\Auni(W_1)$, we have by Eq.~\eqref{eqZpi0AOm} and
the intertwiner relation~\eqref{eqCocInt}   
\begin{align} \label{eqDSD'}
\Delta_\barrho^{it}\, S_\rho\,\Delta_\rho^{-it}\,\pi_0(A)\Om 
&=
\pi_0\big(\barrho(A^*)\coc_\barrho(t)\sigma_t[\barrho(\coc_\rho(-t)^*)R_\rho]
\big)\,\Om. 
\end{align} 
We shall now use a result of Longo~\cite{Longo97}. Namely, we are in
the situation where Propositions 1.3 and 1.4 in~\cite{Longo97} apply, yielding 
\begin{equation*} 
R_\rho^* \barrho\big(\coc_\rho(-t)\big)\coc_\barrho(-t) = \sigma_{-t}
(R_\rho^*).
\end{equation*}
Applying $\sigma_t$, adjoining, and using the cocycle identity
$\coc_\barrho(t)\sigma_t(\coc_\barrho(-t))=1, 
$
see Eq.~\eqref{eqCoc'}, yields 
\begin{equation} \label{eqZZRR}
\coc_\barrho(t)\sigma_t[\barrho(\coc_\rho(-t)^*)R_\rho]=R_\rho.
\end{equation}
Hence Eq.~\eqref{eqDSD'} reads
$$
\Delta_\barrho^{it}\, S_\rho\,\Delta_\rho^{-it}\,\pi_0(A)\Om 
=\pi_0\big(\barrho(A^*)R_\rho\big)\,\Om \equiv S_\rho \,\pi_0(A)\Om.
$$
Since $\Delta_\rho^{it}$ maps the core $\pi_0\Auni(W_1)\Om$ of
$S_\rho$ onto itself by Eq.~\eqref{eqZpi0AOm}, this shows that 
\begin{align*} 
\Delta_\barrho^{it}\, S_\rho\,\Delta_\rho^{-it} &=  S_\rho, 
\end{align*}
which implies Eq.~\eqref{eqDJD} of the Theorem. 
{}For the proof of Eq.~\eqref{eqJUJ} we need the following Lemma. 
\begin{Lem} \label{DSDUSU}
For $a$ in the closure of $W_1$, there holds 
\begin{align} 
U_\barrho(a)^{-1}\, S_\rho\,U_\rho(a) &\subset   S_\rho. \label{eqUSU}
\end{align}
\end{Lem}
\begin{Proof} 
First recall from~\cite{F81,BuF} that the representation 
$\hat U_\barrho$ defined by 
\begin{equation} \label{eqHatUBarrho}
\hat U_\barrho(x)\,\pi_0[\barrho(A)R_\rho]\Om := 
\pi_0[\barrho(\alpha_x(A)Y_\rho(x)^*)R_\rho]\Om 
\end{equation}
implements $\alpha_x$ in the representation $\pi_0\barrho$, i.e.\ 
$\Ad\hat U_\barrho(x)\circ\pi_0\barrho=\pi_0\barrho\circ\alphaO_x$.%
\footnote{We recall the argument in the present setting. The endomorphism
  $\alphaO_{-x}\circ\phi_\rho\circ \beta_x$, where 
$\beta_x:=\Ad Y_\rho(x)\circ \alpha_x$, is a left inverse of $\rho$
  and therefore coincides with $\phi_\rho$ by uniqueness. This implies
  that the state $\varphi_\rho$ is invariant under the automorphism
  group $\beta_x$ and hence that 
$$
U_{\barrho\rho}(x) \pi_0[\barrho(A)R_\rho]\Omega  := 
\pi_0[\barrho\beta_x(A)R_\rho]\Omega
$$
defines a unitary representation of the translations. 
But $\hat U_\barrho(x)$ defined above coincides with 
$\pi_0\barrho(Y_\rho(x)^*) U_{\barrho\rho}(x)$, hence is a
well-defined unitary operator. The implementation property is checked
directly from the definition~\eqref{eqHatUBarrho}, and implies in turn 
the representation property.} 
The representation $\hat U_\barrho$ therefore coincides with $U_\barrho$
up to a one-dimensional representation $c(\cdot)$. 
We now have, for $a$ in the closure of $W_1$ and $A\in\Auni(W_1-a)$, 
\begin{align*} 
S_\rho\,U_\rho(a)\,\pi_0(A)\Om &= 
\pi_0\big[\barrho(\alpha_a(A^*)Y_\rho(a)^*)R_\rho\big]\Om 
= \hat U_\barrho(a)\,\pi_0[\barrho(A^*)R_\rho]\Om \\
& = \hat U_\barrho(a)\,S_\rho\,\pi_0(A)\Om. 
\end{align*}
Since $\hat U_\barrho$ and $U_\barrho$ coincide up to the character $c$
as discussed above, we therefore have 
$$ 
U_\barrho(-a)\,S_\rho \,U_\rho(a)= c(a)\, S_\rho 
\quad \text{ on } \; D:=\calA_0(W_1-a)\Om.
$$
Applying $\Delta_\barrho^{it}\,\cdot\,\Delta_\rho^{-it}$ to this
equation and using the by now established Eq.s~\eqref{eqDUD} and
\eqref{eqDJD} of the Theorem, yields $c(\Boox{t}a)=c(a)$ or 
$c((\unity-\Boox{t})a)=1$. 
By the representation property of $c$, the same holds for $-a\in W_1^\clo$.  
Since $W_1^\clo$ and $-W_1^\clo$ span the whole Minkowski space and 
$1-\Boox{t}$ is invertible for $t\neq0$, this shows that $c$ is
trivial. Since $D$ is a core for the left hand side of relation~\eqref{eqUSU},
this completes the proof. 
\end{Proof}
We are now ready to prove Eq.~\eqref{eqJUJ} of the Theorem. 
To this end, let $a\in W_1^\clo$ and $\phi\in D:=\calA_0(W_1-a)\Om$. 
By Eq~\eqref{eqDUD}, we have for all $t\in\RR$ 
\begin{equation} \label{eqJUJ'} 
\Delta_\rho^{it}\,U_\rho(a) \, \phi
= U_\rho(\Boox{t}a)\,\Delta_\rho^{it} \phi.
\end{equation}
Now by Lemma~\ref{DSDUSU}, the vector
$U_\rho(a) \, \phi$ is in the domain of the 
operator $\Delta_\rho^{1/2}$, hence the left hand side is bounded for
$t$ in the strip $\RR - i[0,1/2]$ and analytic in its
interior. The same holds for the vector valued function 
$t\mapsto \Delta_\rho^{it} \,\phi$ on the right hand side. 
Further, for $a\in W_1^\clo$ the 
operator valued function $t\mapsto U_\rho(\Boox{t}a)$ is norm-bounded on
the strip $\RR - i[0,1/2]$ and analytic in its interior, and at 
$t=-i/2$ has the value $U_\rho(ja)$, see e.g.~\cite[Section V.4.1]{H96}. 
Therefore, Eq.~\eqref{eqJUJ'} implies that 
$$ 
\Delta_\rho^{1/2}\,U_\rho(a) \,\phi 
= U_\rho(ja)\,\Delta_\rho^{1/2} \,\phi. 
$$
Multiplying with $J_\rho$ and using relation~\eqref{eqUSU} of 
Lemma~\ref{DSDUSU} yields 
\begin{equation} \label{eqJUJ''} 
U_\barrho(a) \,S_\rho \,\phi 
= J_\rho U_\rho(ja) J_\rho^{-1}\,S_\rho \,\phi. 
\end{equation}
Since $S_\rho$ has dense range, this shows Eq.~\eqref{eqJUJ} for $x=ja$
in the closure of $jW_1$ and, by adjoining, also for
arbitrary $x$. This completes the proof of Eq.~\eqref{eqJUJ} of the Theorem. 
To prove Eq.~\eqref{eqJJ}, note that Lemma~\ref{SRhoOm} 
implies that $S_\rho=\chi_\rho\,S_\barrho^{-1}$. Using that
$\Delta_\barrho^{-1/2}J_\barrho^{-1}=J_\barrho^{-1}\,\Delta_\rho^{1/2}$ 
by Eq.~\eqref{eqDJD} and that $J_\barrho$ is anti-linear, one gets 
Eq.~\eqref{eqJJ}. This completes the proof of the Theorem. 
\paragraph{Proof of Proposition~\ref{DeltaJImplement}.}  
We now turn to Eq.~\eqref{eqDeltaImplement} of 
Proposition~\ref{DeltaJImplement}. On $\Auni(W_1)$, this equation follows from 
Eq.~\eqref{eqCocInt} by applying $\pi_0$ to the latter. Further, the fact that
$\pi_0^{-1}(\Delta_\rho^{it}\DO^{-it})$ is in $\Auni(W_1)$ and hence
commutes with $\Auni(W_1')$ implies that
$\Ad\Delta_\rho^{it}\circ\pi_0=\pi_0\circ \sigma_t$ on $\Auni(W_1')$. 
Since $\rho$ acts as the identity on $\Auni(W_1')$, this implies
Eq.~\eqref{eqDeltaImplement} on $\Auni(W_1')$. For translates of $W_1$
or $W_1'$, the equation follows from Borchers' commutation relations, 
Eq.s~\eqref{eqDUD} and \eqref{eqSigmaAlphaSigma}.  
Before proving Eq.~\eqref{eqJImplement} of the proposition, we
establish the following 
intertwiner properties of the relative modular conjugation. 
\begin{Lem}[Intertwiner Properties of $\boldsymbol{J_\rho}$.] \label{JInt}
The unitary operators $J_\rho J_0$ and $J_0J_\rho$ 
have the intertwiner properties 
\begin{align} 
 \pi_0\barrho(A) \; J_\rho J_0 & = J_\rho J_0 \; \pi_0(A),  
\label{eqModConjInt} \\ 
\pi_0(A) \; J_0 J_\rho  &= J_0 J_\rho \; \pi_0\rho(A)  \label{eqModConjInt'}
\end{align}
for $A\in \Auni(W_1)$. 
\end{Lem}
\begin{Proof} 
These are consequences of a standard result~\cite{Stratila} which
relates the conjugations of relative Tomita operators, see 
Eq.~\eqref{eqJ21J1} in the Appendix. 
Here, in Eq.~\eqref{eqModConjInt} $S_\rho$ is being considered 
as the relative Tomita operator associated with the pair of states
$(\omega_0,\varphi_\rho)$, characterized by Eq.~\eqref{eqSTomRho}, and 
in Eq.~\eqref{eqModConjInt'} as the relative Tomita operator
associated with the pair $(\varphi_\barrho,\omega_0)$, characterized
by Eq.~\eqref{eqSTomRho'} of Lemma~\ref{SRhoOm}. 
\end{Proof}
We are now ready to prove Eq.~\eqref{eqJImplement} of
Proposition~\ref{DeltaJImplement}.  
By Eq.s~\eqref{eqModConjInt'} and Eq.~\eqref{eqJ0alphaj} 
we have on $\Auni(W_1)$ 
$$
\Ad J_\rho \circ \pi_0\rho 
\equiv \Ad J_0 \circ\Ad(J_0J_\rho)\circ \pi_0\rho  
=\pi_0\circ \alphaO_j =\pi_0\barrho\circ \alphaO_j, 
$$ 
since $\barrho$ acts as the identity on 
$\alphaO_j\Auni(W_1)\equiv\Auni(W_1')$, while by 
Eq.~\eqref{eqModConjInt} and Eq.~\eqref{eqJ0alphaj} we have on $\Auni(W_1')$ 
$$
\Ad J_\rho \circ \pi_0\rho =\Ad J_\rho \circ \pi_0 
\equiv \Ad (J_\rho J_0)\circ\Ad J_0\circ\pi_0
=\pi_0\barrho\circ \alphaO_j.
$$ 
This shows that Eq.~\eqref{eqJImplement} holds on
$\Auni(W_1)\cup\Auni(W_1')$. Borchers' commutation relations then
imply that it holds on $\Auni(W)$, $W\in\calW_1$, completing the proof
of Proposition~\ref{DeltaJImplement}. 
\appendix
\renewcommand{\theequation}{\thesection.\arabic{equation}}
\setcounter{equation}{0}
\renewcommand{\theLem}{\thesection.\arabic{Lem}}
\section{Relative Tomita Operators} \label{secRelMod}
We recall the relevant notions from relative Tomita theory, 
following~\cite{Stratila}. (For the standard Tomita theory,
see e.g.~\cite{BraRob} and Eq.~\eqref{eqSTomO} above.) 
Let $\calM$ be a von Neumann
algebra and $\varphi_1$, $\varphi_2$ two faithful normal positive
functionals on $\calM$, and denote by $\sigma_t^1$ and $\sigma_t^2$ the
respective modular automorphism groups. 
Then there exists a family of unitaries
$Z_{21}(t)\in \calM$ satisfying the intertwiner and cocycle properties 
\begin{align} \nonumber 
\sigma_t^2(A)\, Z_{21}(t)&= Z_{21}(t)\, \sigma_t^1(A) ,\\
Z_{21}(t+s)&=Z_{21}(t)\sigma_t^1(Z_{21}(s)), \label{eqCoc'}
\end{align}
respectively, and characterized by a certain KMS property. These facts
have been shown by Connes~\cite{Connes73} and are reviewed
in~\cite[Sect.~I.3.1]{Stratila}. 
The family $Z_{21}(t)$ is called the
Connes-cocycle associated with  the pair $\varphi_1$ and $\varphi_2$
and usually denoted by $(D\varphi_1:D\varphi_2)_t$.  
This cocycle may be expressed in terms of the corresponding GNS
representations as follows~\cite[Sect.~I.3.11]{Stratila}. 
Let $(\calH_i,\pi_i,\xi_i)$ be the GNS triples of $\varphi_i$,
$i=1,2$. Then the operator $S_{21}$ from $\calH_1$ to $\calH_2$ defined by 
\begin{equation*} 
S_{21}\,\pi_1(A)\xi_1 := \pi_2(A^*)\xi_2, \quad A\in \calM, 
\end{equation*}
is closable. We denote its closure by the same symbol, and its polar
decomposition by 
$$ 
S_{21}= J_{21} \, \Delta_{21}^{1/2}. 
$$
These operators are called the relative Tomita modular objects 
associated with the pair $\varphi_1$ and $\varphi_2$.  
Let now $\Delta_1^{it}$ denote the unitary modular group of
$\pi_1(\calM)$ and $\xi_1$. Then $\Delta_{21}^{it}\Delta_1^{-it}$ is 
in $\pi_1(\calM)$ and coincides with $\pi_1(Z_{21}(t))$, i.e.\ there 
holds~\cite[Sect.~I.3.11]{Stratila}  
\begin{equation} \label{eqConnesCoc}
Z_{21}(t) = \pi_1^{-1}\big(\Delta_{21}^{it}\Delta_1^{-it}\big).
\end{equation}
Finally, as shown in~\cite[Sect.~I.3.16]{Stratila}, the unitary operator 
$$
V_{21} := J_{21}\,J_1 \equiv J_2\,J_{21}, 
$$
where $J_i$ is the modular conjugation of $\pi_i(\calM)$ and $\xi_i$,
$i=1,2$, is an intertwiner from $\pi_1$ to $\pi_2$, that means it satisfies 
\begin{equation} \label{eqJ21J1}
\pi_2(A)\, V_{21} = V_{21}\, \pi_1(A),\quad A\in \calM. 
\end{equation}
\paragraph{Acknowledgements.}
I gratefully acknowledge financial support by 
FAPEMIG.
\providecommand{\bysame}{\leavevmode\hbox to3em{\hrulefill}\thinspace}
\providecommand{\MR}{\relax\ifhmode\unskip\space\fi MR }
\providecommand{\MRhref}[2]{%
  \href{http://www.ams.org/mathscinet-getitem?mr=#1}{#2}
}
\providecommand{\href}[2]{#2}

\end{document}